\documentclass[twocolumn,showpacs,preprintnumbers,amsmath,amssymb,floatfix]{revtex4}

\usepackage{graphicx}% Include figure files
\usepackage{dcolumn}% Align table columns on decimal point
\usepackage{bm}% bold math

\begin{document}

%%%%
%    Greek Letters
%

\let\a=\alpha      \let\b=\beta       \let\c=\chi        \let\d=\delta
\let\e=\varepsilon \let\f=\varphi     \let\g=\gamma      \let\h=\eta
\let\k=\kappa      \let\l=\lambda     \let\m=\mu
\let\o=\omega      \let\r=\varrho     \let\s=\sigma
\let\t=\tau        \let\th=\vartheta  \let\y=\upsilon    \let\x=\xi
\let\z=\zeta       \let\io=\iota      \let\vp=\varpi     \let\ro=\rho
\let\ph=\phi       \let\ep=\epsilon   \let\te=\theta
\let\n=\nu
\let\D=\Delta   \let\F=\Phi    \let\G=\Gamma  \let\L=\Lambda
\let\O=\Omega   \let\P=\Pi     \let\Ps=\Psi   \let\Si=\Sigma
\let\Th=\Theta  \let\X=\Xi     \let\Y=\Upsilon

%
%%%

%%%
%    Calligraphic letters
%

\def\cA{{\cal A}}                \def\cB{{\cal B}}
\def\cC{{\cal C}}                \def\cD{{\cal D}}
\def\cE{{\cal E}}                \def\cF{{\cal F}}
\def\cG{{\cal G}}                \def\cH{{\cal H}}
\def\cI{{\cal I}}                \def\cJ{{\cal J}}
\def\cK{{\cal K}}                \def\cL{{\cal L}}
\def\cM{{\cal M}}                \def\cN{{\cal N}}
\def\cO{{\cal O}}                \def\cP{{\cal P}}
\def\cQ{{\cal Q}}                \def\cR{{\cal R}}
\def\cS{{\cal S}}                \def\cT{{\cal T}}
\def\cU{{\cal U}}                \def\cV{{\cal V}}
\def\cW{{\cal W}}                \def\cX{{\cal X}}
\def\cY{{\cal Y}}                \def\cZ{{\cal Z}}
%
%%%%

\newcommand{\Ns}{N\hspace{-4.7mm}\not\hspace{2.7mm}}
\newcommand{\qs}{q\hspace{-3.7mm}\not\hspace{3.4mm}}
\newcommand{\ps}{p\hspace{-3.3mm}\not\hspace{1.2mm}}
\newcommand{\ks}{k\hspace{-3.3mm}\not\hspace{1.2mm}}
\newcommand{\des}{\partial\hspace{-4.mm}\not\hspace{2.5mm}}
\newcommand{\desco}{D\hspace{-4mm}\not\hspace{2mm}}

%%%%

%\draft command makes pacs numbers print
%\draft

\title{$B \rightarrow J/\psi\, K^*$ in a Supersymmetric Right-handed
Flavor Mixing Scenario}
\author{Wei-Shu Hou}
%\email{wshou@phys.ntu.edu.tw}
\author{Makiko Nagashima}
%\email{}
\author{Andrea Soddu}
%\email{asoddu@hep1.phys.ntu.edu.tw}
\affiliation{Department of Physics, National Taiwan University, Taipei, Taiwan 106, R.O.C.}
\date{\today}

\begin{abstract}
A supersymmetric extension of the Standard Model, with maximal
$\tilde s_R$-$\tilde b_R$ mixing and a new source of $CP$ violation,
contain all the necessary ingredients to account for a possible
anomaly in the measured $CP$ asymmetry in $B\rightarrow \phi K_S$
decay. In the same framework we study the decay $B\rightarrow
J/\psi\, K^*$, paying particular attention to observables that can
be extracted by performing a time dependent angular analysis, and
become nonzero because of new physics effects.
\end{abstract}

\maketitle
%\narrowtext

\section{Introduction}

Recent results on $B$ decay to charmonium modes, such as ${B
\rightarrow J/\psi\, K_S (K^*)}$, are in good agreement with the
Standard Model (SM). However, for the time-dependent $CP$
asymmetry in ${B \rightarrow \phi K_S}$ mode, also if new results
seem to show more agreement with the SM, $S_{\phi K_S}=0.50 \pm 0.25 $
(BaBar) \cite{BaBarphiKs} and $S_{\phi K_S}=0.06 \pm 0.33$
(Belle) \cite{BellephiKs}, than in the past,
(before ICHEP04, the world average
$S_{\phi K_S}=0.02 \pm 0.29 $ \cite{Moriond}), New Physics (NP)
effects could still play some role. A deviation from the SM
expectation of $S_{\phi K_S} \cong S_{\psi K_S}$ would call for
large $s-b$ mixing, the existence of a new source of $CP$
violation, and perhaps right-handed dynamics
\cite{ChuaHouNagashima}. It would be important to find confirming
evidence in the future.

In anticipation of a future ``super B factory" that would allow
precision measurements, we study $CP$ violation in the
vector-vector ${B_d^0 \rightarrow J/\psi\, K^*}$ mode by taking
into account deviations in $S_{\phi K_S}$. We do not expect the NP
effects to show up in the branching fraction since, in contrast to
$B \rightarrow \phi K_S$, ${B_d^0 \rightarrow J/\psi K^*}$ decay
is tree dominant. With special attention to observables related to
the time dependent $CP$ asymmetry \cite{HeHou1,Fleischer,LondonSinha}, we
focus on manifestations of NP effects. In the phenomenological
analysis presented in this paper, we pay particular attention to
those obserables which are expected to vanish in the SM
\cite{LondonSinha}.

\section{$B \rightarrow V V$ }

The ${B_d^0 \rightarrow J/\psi K^{(*)}}$ decay is dominated by the
tree level $\bar{b}\to c\bar c \bar{s}$ process, while the $CP$
phase in the corresponding penguin amplitude is highly suppressed.
If NP contributions are present, it could manifest itself as
direct $CP$ violation effects. This can be illustrated by the full
amplitude for the decay $B \rightarrow f$,
\begin{equation}
A(B\rightarrow f) = a e^{i\d^a}+be^{i\phi}e^{i\d^b} \, ,
\label{twoAmp}
\end{equation}
where the weak phases are assumed to be zero for the first
amplitude and $\phi$ for the second, and $\d^{a,b}$ are the
respectively strong phases. For the $CP$ conjugate decay $\bar{B}
\rightarrow \bar{f}$ the amplitude is given by changing the sign
of the weak phase $\phi$. One then obtains the direct $CP$
asymmetry
\begin{eqnarray}
a_{\rm dir}^{CP}
 & = &\frac{\G(B\rightarrow f)-\G(\bar{B} \rightarrow \bar{f})}
           {\G(B\rightarrow f)+\G(\bar{B} \rightarrow \bar{f})} \\
 & = &\frac{2ab\sin{(\d^a-\d^b)}\sin{\phi}}
           {a^2+b^2+2ab\cos{(\d^a-\d^b)}\cos{\phi}} \, ,
\label{dirCPasymm}
\end{eqnarray}
which does not vanish for $\phi\neq 0$ and if the strong phase
difference is also not zero. However, since the $b$ quark is
rather heavy, the strong phases are expected to be small and the
effect of NP in the direct $CP$ asymmetry could be washed out. It
is therefore important that one can still seek for NP effects by
performing a time dependent analysis and comparing $B_d^0(t)
\rightarrow J/\psi K^*$ with $J/\psi K_S$. In fact, more
information is contained in the time dependent angular analysis of
vector-vector decays such as $B_d^0(t) \rightarrow J/\psi K^*$ or
$\phi K^*$ \cite{Fleischer,LondonSinha}.

For a $B \rightarrow VV$ decay, the final state can be decomposed
into three helicity amplitudes $\{A_0,A_{\parallel},A_{\perp}\}$.
$A_0$ corresponds to both the vector mesons being polarized along
their direction of motion, while $A_{\parallel}$ and $A_\perp$
correspond to both polarization states being transverse to their
directions of motion but parallel and orthogonal to each other,
respectively \cite{Rosner}. If in particular we consider the decay
$B_d^0(t) \rightarrow J/\psi K^*$, analogous to Eq.~(\ref{twoAmp})
we have,
\begin{eqnarray}
A_{\l}(B\rightarrow J/ \psi K^*) & = & a_{\l}
e^{i\d_{\l}^a}+b_{\l}e^{i\phi}e^{i\d_{\l}^b} , \label{Amplambda1}
\\
\bar{A}_{\l}(\bar{B}\rightarrow J/ \psi \bar{K}^*) & = & a_{\l}
e^{i\d_{\l}^a}+b_{\l}e^{-i\phi}e^{i\d_{\l}^b}, \label{Amplambda2}
\end{eqnarray}
where $a_{\l}$ and $b_{\l}$ are the SM and NP amplitudes and
$\d_{\l}^{a,b}$ their respective strong phases, for each helicity
component. The full decay amplitude becomes
\begin{eqnarray}
A(B\rightarrow J/ \psi K^*)
 & = & A_0g_0+A_{\parallel}g_{\parallel}+iA_{\perp}g_{\perp}, \\
\bar A(\bar{B}\rightarrow J/ \psi \bar{K}^*)
 & = & \bar{A}_0g_0+\bar{A}_{\parallel}g_{\parallel}-i\bar{A}_{\perp}g_{\perp},
\end{eqnarray}
with $g_{\l}$ the coefficients of the helicity amplitudes in the
linear polarization basis \cite{Sinha}. If one considers the case
where $K^*$ and $\bar{K}^*$ are detected through their decay to
$K_S \pi^0$ so that both $B_d^0$ and $\bar{B}_d^0$ decay to a
common final state, the time dependent decay rates can be written
as,
\begin{widetext}
\begin{equation}
\G(B_d(\bar{B}_d)\rightarrow J/\psi\, K^*) =
 e^{-\G t}\sum_{\l\leq\s} \left(\L_{\l\s} \pm \Si_{\l\s}\cos\D mt
  \mp \r_{\l\s}\sin\D mt \right) \, g_{\l}g_{\s} \, .
\label{rate}
\end{equation}
\end{widetext}

By performing an angular analysis and time dependent study of the
decays $\bar{B}_d \rightarrow J/ \psi \bar{K}^*$ and $B_d
\rightarrow J/ \psi K^*$, one can measure the observables $\L_{\l
\s}$, $\Si_{\l \s}$ and $\r_{\l \s}$ \cite{LondonSinha}. These
observables can be expressed in terms of the normalized helicity
amplitudes $A_0$, $A_{\parallel}$ and $A_{\perp}$:

%\begin{widetext}
\begin{eqnarray}
\L_{\l \l}& = &\frac{{\mid A_{\l} \mid}^2 + {\mid \bar{A}_{\l} \mid}^2}{2} \, ,
 \nonumber \\
\Si_{\l \l} & = &\frac{{\mid A_{\l} \mid}^2 - {\mid \bar{A}_{\l} \mid}^2}{2} \, ,
 \nonumber \\
\L_{\perp i}& = &-{\rm Im}\,(A_{\perp}A_i^*-\bar{A}_{\perp}\bar{A}_i^*) \, ,
 \nonumber \\
\L_{\parallel 0}& = &{\rm Re}\,(A_{\parallel}A_0^*+\bar{A}_{\parallel}\bar{A}_0^*) \, ,
 \nonumber \\
\Si_{\perp i} & = &-{\rm Im}\,(A_{\perp}A_i^*+\bar{A}_{\perp}\bar{A}_i^*) \, ,
 \nonumber \\
\Si_{\parallel 0}& = &{\rm Re}\,(A_{\parallel}A_0^*-\bar{A}_{\parallel}\bar{A}_0^*) \, ,
 \nonumber \\
\r_{\perp i} & = &-{\rm Re}\left(\frac{q}{p}
 (A_{\perp}^*\bar{A}_i+A_i^*\bar{A}_{\perp})\right) \, ,
 \nonumber \\
\r_{\perp \perp} & = &-{\rm Im}\left(\frac{q}{p}A_{\perp}^*\bar{A_{\perp}} \right) \, ,
 \nonumber \\
\r_{\perp \perp} & = &{\rm Im}\left(\frac{q}{p}
 (A_{\parallel}^*\bar{A}_0+A_0^*\bar{A}_{\parallel}) \right) \, ,
 \nonumber \\
\r_{i i}& = &{\rm Im}\left(\frac{q}{p}A_{i}^*\bar{A_{i}} \right)
\, , \label{Observ}
\end{eqnarray}
%\end{widetext}
%
where $i=\{0,\parallel\}$, $q/p=\exp\,(-2i\phi_{\rm mix})$ with
$\phi_{\rm mix}$ the weak phase in $B_d^0-\bar{B}_d^0$ mixing.
From Eqs. (\ref{Amplambda1}) and (\ref{Amplambda2}) one can obtain
the same observables in terms of $a_{\l}$, $b_{\l}$, $\phi$,
$\d_{\l}\equiv \d_{\l}^{b}-\d_{\l}^{a}$ and $\D_i \equiv
\d_{\perp}^{b}-\d_{i}^{a}$ \cite{LondonSinha}. In particular,
$\L_{\perp i}$ can be expressed as,
\begin{equation}
\L_{\perp i} = 2[a_{\perp}b_{i}\cos(\D_i-\d_i)-a_i
b_{\perp}\cos(\D_i+\d_{\perp})]\sin\phi \, .
\label{lambdaperpi}
\end{equation}
The observable $\L_{\perp i}$ is special, as made clear by
Eq.~(\ref{lambdaperpi}), because it remains nonzero in the
presence of NP effects ($\phi \neq 0$), even if the strong phase
differences vanish. In contrast, direct $CP$ asymmetries
$\Si_{\l\l}$ are washed out if the strong phase differences vanish
\cite{LondonSinha}.

\section{$B \rightarrow J/\psi\, K^*$}

We can now proceed towards NP effects in $B \rightarrow J/\psi\,
K^*$. We start by writing the decay amplitudes for $B \rightarrow
J/\psi\, K^*$ using the factorization approximation, but keeping
the color octet contribution \cite{HeHou2},
\begin{widetext}
\begin{eqnarray}
A_{\l}(B \rightarrow  J/\psi K^*) & = &
i\frac{G_F}{\sqrt{2}}V_{cb}V_{cs}^*f_{\psi}m_{\psi}\e_{\psi}^{\mu}(\l)\{
a_2^{\rm eff} \langle K^*\mid \bar{s}\g_{\mu}(1-\g_5)b \mid B \rangle
\nonumber \\
& - &\frac{\a_s}{2 \pi}\frac{m_b}{q^2} \x_8^{\prime} \langle
K^*\mid \bar{s}i\s_{\mu\nu}q^{\nu}
(c_{12}\g_{\mu}(1+\g_5)+c_{12}^{\prime}(1-\g_5))b\mid B \rangle \}
\, , \label{AmpBtoJPsiKs}
\end{eqnarray}
\end{widetext}
where $\l=\l_{J/\psi}=\l_{K^*}=0,\ \pm1$ denote the helicities of
the final state vector particles $J/\psi$ and $K^*$ in the $B^0$
rest frame \cite{KramerPalmer}. The dominant contribution in
Eq.~(\ref{AmpBtoJPsiKs}) is given by the tree level term
proportional to
\begin{equation}
a_2^{\rm eff} = c_2^{\rm eff} +  \z \,c_1^{\rm eff} \, ,
\label{a2eff}
\end{equation}
while the color dipole moment terms, with the $c_{12}$ operator
coming dominantly from SM and the $c_{12}^{\prime}$ operator due
exclusively to NP, give smaller corrections. In the expression for
$a_2^{\rm eff}$ we have neglected the strong and electroweak
penguin contributions. The quantity $\z=\frac{1}{N_c} + \x_8$ in
Eq.~(\ref{a2eff}) takes the value  $1/N_c=1/3$ in the naive
factorization while deviations from $1/N_c$ due to
non-factorizable contributions to the hadronic matrix elements are
measured by the parameters $\x_8$ and $\x_8^{\prime}$
\cite{HeHou2}.
%
%For the definition of the hadronic parameters $\x_8$ and $\x_8^{\prime}$ we follow Ref. \cite{NeubertStech}
%
%\begin{eqnarray}
%\x_8 & = & \frac{\langle J/\psi K^* \mid [\bar{c}\g_{\mu}\l^a c]\,[ \bar{s}\g^{\mu}(1-g_5)\l^a b] \mid B \rangle}
%{2 \langle J/\psi \mid \bar{c} \g^{\mu} c\mid \rangle\langle K^*\mid \bar{s}\g^{\mu}(1-\g_5)b \mid B \rangle}
%\label{csi8} \, ,\\
%\x_8^{\prime} & = & \frac{\langle J/\psi K^* \mid [\bar{c}\g_{\mu}\l^a c]\,
%[ \bar{s}i\s^{\mu\nu}q_{\nu}(1\pm g_5)\l^a b] \mid B \rangle}
%{2\langle J/\psi \mid \bar{c} \g^{\mu} c\mid \rangle\langle K^*\mid \bar{s}i\s^{\mu\nu}(1\pm\g_5)b \mid B \rangle}
%\label{csi8prime}
%\end{eqnarray}
%
The effective Wilson coefficients $c_1^{\rm eff}(m_b)$ and
$c_2^{\rm eff}(m_b)$ for a $b \rightarrow s$ transition are
defined in Ref. \cite{CCTY}.

The way we proceed to determine the parameter $\z$ follows Ref.
\cite{NeubertStech}. We fit the branching ratios for the decays $B
\rightarrow J/\psi\, K_S(K^*)$ and $B \rightarrow \psi(2S)\, K^*$
to extract $a_2^{\rm eff}$.
%, as indicated in Figs. 1-3.
%
We checked explicitly that the NP effect of
Ref.~\cite{ChuaHouNagashima} does not make significant impact on
the decay rates.
However, the extraction of $a_2^{\rm eff}$ depends on the specific
model one uses for the hadronic form factors \cite{NeubertStech}.
In this work we use the form factors at zero momentum transfer for
the $B\rightarrow V$ transitions obtained in the light-cone sum
rule (LCSR) analysis \cite{BallBraun}. The form factor  $q^2$
dependence is parametrized by
\begin{equation}
f(q^2)=\frac{f(0)}{1-a(q^2/m_B^2)+b(q^2/m_B^2)^2} \, ,
\label{fqsquare}
\end{equation}
where the values of the parameters $a$ and $b$ are given in
Ref~\cite{BallBraun}.
%
%\begin{figure}[!ht]
%\scalebox{0.42}{\includegraphics{Figure_1.eps}}
%\caption{$BR(B \rightarrow J/\Psi K_S)$ as a function of $\s$ for
%$\tilde{m}_1=200 {\rm GeV}$, $\tilde{m}=2 {\rm TeV}$ and
%$m_{\tilde{g}}=300 {\rm GeV}$ compared with experiment.
%\label{Fig_1}}
%\end{figure}
%\begin{figure}[!ht]
%\scalebox{0.42}{\includegraphics{Figure_2.eps}} \caption{$BR(B
%\rightarrow J/\Psi K^*)$ as a function of $\s$ for
%$\tilde{m}_1=200 {\rm GeV}$, $\tilde{m}=2 {\rm TeV}$ and
%$m_{\tilde{g}}=300 {\rm GeV}$  compared with experiment.
%\label{Fig_2}}
%\end{figure}
%\begin{figure}[!ht]
%\scalebox{0.42}{\includegraphics{Figure_3.eps}} \caption{$BR(B
%\rightarrow \Psi(2s) K^*)$ as a function of $\s$ for
%$\tilde{m}_1=200 {\rm GeV}$, $\tilde{m}=2 {\rm TeV}$ and
%$m_{\tilde{g}}=300 {\rm GeV}$ compared with experiment.
%\label{Fig_3}}
%\end{figure}
%
Our extracted value for $a_2^{\rm eff}$ is $0.20$, which using
Eq.~(\ref{a2eff}) and the effective Wilson coefficients of Table 1
of Ref.~\cite{CCTY} gives $\z=0.48$, or the effective number of
colors $1/\z=2.1$. Knowing $\z$ one derives $\x_8=0.15$ and for
the value of $\x_8^{\prime}$ we assume $\x_8^{\prime}=\x_8$
\cite{HeHou2}.

%By looking at the behavior of the branching ratios for $B
%\rightarrow J/\psi K^*$ versus the NP weak phase $\sigma$, one can
%observe that NP contributions are of the order of a few percent.
%One may therefore expect that the observables which can be
%measured by performing an angular analysis will not differ from
%the SM predictions by more than a few percent.

\section{$B \rightarrow J/\psi K^*$ and  NP Effects}

As previously mentioned, in this work we focus on the decay
$B_d^0(t) \rightarrow J/\psi\, K^*$ in the context of a
supersymmetric model with maximal $\tilde{s}_R-\tilde{b}_R$
mixing. An approximate Abelian flavor symmetry \cite{LNS} can be
introduced to justify such a large mixing. In this model the
right-right mass matrix $\tilde{M}^{2(sb)}_{RR}$ for the
strange-beauty squark sector takes the form
\begin{equation}
\tilde{M}^{2(sb)}_{RR} = \left[\begin{array}{cc}
\tilde{m}_{22}^2  &  \tilde{m}_{23}^2e^{-i\s} \\
\tilde{m}_{23}^2e^{i\s} & \tilde{m}_{33}^2
\end{array} \right] \equiv
R\left[\begin{array}{cc}
\tilde{m}_{1}^2  &  0 \\
0 & \tilde{m}_{2}^2
\end{array} \right]R^{\dag},
\label{MRR}
\end{equation}
where $\tilde{m}_{ij} \sim \tilde{m}^2$, the squark mass scale,
and
\begin{equation}
R=\left[\begin{array}{cc}
\cos{\theta}  & \sin{\theta}   \\
-\sin{\theta}e^{i\s} & \cos{\theta}e^{i\s}
\end{array} \right]\, .
\end{equation}
The phase $\sigma$ is the NP weak phase which will affect the
observables one can extract from an angular analysis.
Because of the almost democratic structure of
$\tilde{M}^{2(sb)}_{RR}$, one of the two strange-beauty squarks,
$\tilde{sb}_1$, can be rather light, even for $\tilde{m}\sim
O({\rm TeV})$. A light strange-beauty squark, together with a
light gluino, can make $S_{\phi,K_s}$ negative for $\sigma
\lesssim \pi/2$ \cite{ChuaHouNagashima}.

\begin{figure}[t!]
\scalebox{0.65}{\includegraphics{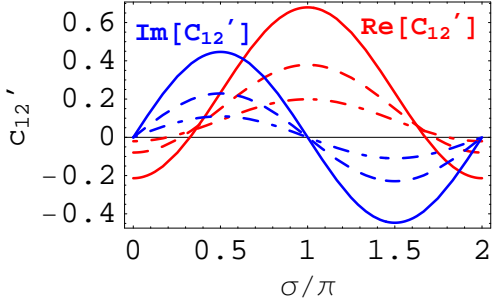}}
\caption{$c_{12}^{\prime}$ as a function of $\s$ for
$\tilde{m}_1=200\ {\rm GeV}$ and $\tilde{m}=2\ {\rm TeV}$. ${\rm
Re}\,(c_{12}^{\prime})$ and ${\rm Im}\,(c_{12}^{\prime})$ are
plotted with solid, dash and solid-dash lines for
$m_{\tilde{g}}=300,\ 500,\ 800\ {\rm GeV}$. \label{Fig_1}}
\end{figure}

The main new contribution to $B_d^0(t) \rightarrow J/\psi\, K^*$
is given by the color dipole moment amplitude through gluino and
$\tilde{sb}$ squark exchange in the loop. The analytic expressions
for the Wilson coefficient $c_{12}^{\prime}$ for the color dipole
moment operator $g_s/(8\pi^2)m_b\bar{s}_{\a}\s^{\mu\nu}(1\pm
g_5)\l^A_{\a\b}/2 \,b_{\b}G^A_{\mu\nu}$ can be found in
Ref.~\cite{ArhribChuaHou}. The coefficient $c_{12}$ remains
basically the same as $c_{12}^{\rm SM}=-0.15$. In Fig.~1 we plot
the real and imaginary parts of the Wilson coefficient
$c_{12}^{\prime}$ for three different values of gluino mass
$m_{\tilde{g}}=300,\ 500,\ 800$ GeV \footnote{A light gluino of
$300$ GeV is less favored by the $b\rightarrow s \gamma$
constraint.}. The eigenvalues of Eq.~(\ref{MRR}) are taken (with
some level of tuning) as $\tilde{m}_1^2=(200\ {\rm GeV})^2$ and
$\tilde{m}_2^2=2\tilde{m}^2-\tilde{m}_1^2$ with $\tilde{m}^2=(2\
{\rm TeV})^2$.

Using the parametrization for the matrix elements $\langle K^*\mid
\bar{s}\g_{\mu}(1-\g_5)b \mid B \rangle$ and $\langle K^*\mid
\bar{s}i\s_{\mu\nu}\g_{\mu}(1\pm\g_5)b\mid B \rangle$ given in
Ref.~\cite{HeHou2} the amplitudes $A_{\l}(B \rightarrow  J/\psi
K^*)$ can be written as,
%
%\begin{widetext}
%\begin{eqnarray}
%A_{\l}(B \rightarrow  J/\psi K^*)
% & = & i\frac{G_F}{\sqrt{2}}V_{cb}V_{cs}^*f_{\psi}m_{\psi} \,
%  \ep^*_{K^*\m}(\l)\ep^*_{\psi\n}(\l)
%  \left\{a_2^{\rm eff}\left[\frac{V(m_{\psi}^2)}{m_B+m_{K^*}}\e^{\m\n\a\b}
%  \,p_{K^* \a} p_{B \b} \right. \right.
%  \nonumber \\
% & - & \left. \left. \frac{i}{2}(m_B+m_{K^*})A_1(m_{\psi}^2)\, g^{\m\n}
%  + i\frac{A_2(m_{\psi}^2)}{m_B+m_{K^*}}\, p_B^{\m}p_B^{\n} \right]
%  \right. \nonumber \\
% & - & \left. \frac{\a_s}{2\pi}\frac{m_B}{m_{\psi}^2}\x_8(c_{12}+c_{12}^{\prime})
%  \, g_{+}(m_{\psi}^2)\e^{\m\n\a\b}\, p_{K^* \a}p_{B \b} \right.
%  \nonumber \\
% & + & \left. i \frac{\a_s}{2\pi}\frac{m_B}{m_{\psi}^2}\x_8(c_{12}-c_{12}^{\prime})
%  \left[\frac{1}{2}(g_{+}(m_{\psi}^2)(m_B^2-m_{K^*}^2)+g_{+}(m_{\psi}^2)m_{\psi}^2)
%  \, g^{\m \n} \right.\right.
%  \nonumber \\
% & - & \left. \left. (g_{+}(m_{\psi}^2)-h(m_{\psi}^2)m_{\psi}^2)
%  \, p_{B \m} p_{B \n} \right] \right\} \, ,
%\label{AlambdaHou}
%\end{eqnarray}
\begin{widetext}
\begin{eqnarray}
A_{\l}(B \rightarrow  J/\psi K^*)
 & = & i\frac{G_F}{\sqrt{2}}V_{cb}V_{cs}^*f_{\psi}m_{\psi} \,
  \ep^*_{K^*\m}(\l)\ep^*_{\psi\n}(\l)
  \left\{a_2^{\rm eff}\left[\frac{V(m_{\psi}^2)}{m_B+m_{K^*}}\e^{\m\n\a\b}
  \,p_{K^* \a} p_{B \b}
  -  \frac{i}{2}(m_B+m_{K^*})A_1(m_{\psi}^2)\,
 g^{\m\n} \right. \right.
  \nonumber \\
 & + & \left. \left. i\frac{A_2(m_{\psi}^2)}{m_B+m_{K^*}}\, p_B^{\m}p_B^{\n} \right]
 - \frac{\a_s}{2\pi}\frac{m_B}{m_{\psi}^2}\x_8(c_{12}+c_{12}^{\prime})
  \, g_{+}(m_{\psi}^2)\e^{\m\n\a\b}\, p_{K^* \a}p_{B \b}
 +  i \frac{\a_s}{2\pi}\frac{m_B}{m_{\psi}^2}\x_8(c_{12}-c_{12}^{\prime})
\right. \nonumber
 \\
  & \times & \left. \left[\frac{1}{2}(g_{+}(m_{\psi}^2)(m_B^2-m_{K^*}^2)+g_{+}(m_{\psi}^2)m_{\psi}^2)
  \, g^{\m \n}
  - (g_{+}(m_{\psi}^2)-h(m_{\psi}^2)m_{\psi}^2)
  \, p_{B \m} p_{B \n} \right] \right\} \, ,
\label{AlambdaHou}
\end{eqnarray}

The general covariant form for $A_{0,\pm 1}$ is given by,
%
%\begin{widetext}
\begin{equation}
A_{0,\pm 1} = \ep^*_{\psi\mu}(0,\pm 1)\ep^*_{K^*\nu}(0,\pm 1)
\left[a g^{\mu\nu} + \frac{b}{m_{\psi}m_{K^*}}
\, p_B^{\mu}p_B^{\nu}+\frac{ic}{m_{\psi}m_{K^*}}
\ep^{\mu\nu\a\b}p_{K^* \a}p_{B \b} \right] \, ,
\label{Alambda}
\end{equation}
%\end{widetext}
%
where $a$, $b$ and $c$ are three invariant amplitudes. The
corresponding amplitudes $\bar{A}_{0,\pm1}$ are obtained by taking
the conjugate of the invariant amplitudes $a,\ b,\ c$ and
switching the sign of the term $\ep^{\mu\nu\a\b}\, p_{K^* \a}p_{B
\b}$ in Eq.~(\ref{Alambda}). By comparing Eq.~(\ref{AlambdaHou})
with Eq.~(\ref{Alambda}) one can extract the three invariant
amplitudes,
%
%\begin{widetext}
\begin{eqnarray}
a & = & \frac{1}{2}(m_B+m_{K^*})A_1(m_{\psi}^2) a_2^{\rm eff}
 - \frac{\a_s}{2 \pi}\frac{m_B}{m_{\psi}^2}\x_8(c_{12}-c_{12}^{\prime})
  \frac{1}{2}\left(g_{+}(m_{\psi}^2)(m_B^2-m_{K^*}^2)
 + g_{+}(m_{\psi}^2)m_{\psi}^2\right) \, ,
 \label{aamp} \\
b & = & -m_{\psi}m_{K^*}\left[\frac{A_2(m_{\psi}^2)}{m_B+m_{K^*}}a_2^{\rm eff}
 + \frac{\a_s}{2 \pi}\frac{m_B}{m_{\psi}^2}\x_8(c_{12}-c_{12}^{\prime})
  \left(g_{+}(m_{\psi}^2)-h(m_{\psi}^2)m_{\psi}^2\right)\right] \, ,
 \label{bamp} \\
c & = & m_{\psi}m_{K^*}\left[\frac{V(m_{\psi}^2)}{m_B+m_{K^*}}a_2^{\rm eff}
 - \frac{\a_s}{2 \pi}\frac{m_B}{m_{\psi}^2}\x_8(c_{12}+c_{12}^{\prime})
  g_{+}(m_{\psi}^2)\right] \, .
 \label{camp}
\end{eqnarray}
\end{widetext}

In the expressions of Eqs. (\ref{aamp})-(\ref{camp}) we have
omitted the common factor $\sqrt{2}G_F
V_{cb}V_{cs}^*f_{\psi}m_{\psi}$. Note that each invariant
amplitude contains a SM contribution which is dominated by the
tree level term proportional to $a_2^{\rm eff}$ plus the color
dipole moment term proportional to $c_{12}$, and a NP contribution
proportional to $c_{12}^{\prime}$. One can consequently write the
three invariant amplitudes as the sum of a SM and a NP
contribution: $a,\ b,\ c \equiv (a,\ b,\ c)^{\rm SM} + (a,\ b,\
c)^{\rm NP}$.

We now rewrite the helicity amplitudes $A_{0, \pm 1}$ in terms of
the invariant amplitudes and the kinematic factor $x \equiv
p_{\psi}\cdot p_{K^*}/(m_{\psi}m_{K^*})$ \cite{DigheDL}. Writing
separately for the SM and NP contributions, we obtain,
\begin{eqnarray}
A_{\pm 1}^{\rm SM,NP} & = & a^{\rm SM,NP} \pm c^{\rm SM,NP}\sqrt{x^2-1} \ , \nonumber \\
A_0^{\rm SM,NP} & = & -a^{\rm SM,NP}x-b^{\rm SM,NP}(x^2-1) \, ,
\end{eqnarray}
with $A_{0, \pm 1}=A_{0,\pm 1}^{SM}+A_{0, \pm 1}^{NP}$.
To evaluate the observables of Eq. (\ref{Observ}) one can transform $A_{0,\pm 1}^{\rm SM,NP}$ in the corresponding
linear polarization amplitudes $A_{0,\parallel,\perp}^{\rm SM,NP}$ using the relations:
${A_{\parallel,\perp}=(A_{+1}\pm A_{-1})/\sqrt{2}}$,
$A_{0}$ being the same in both basis.

The invariant amplitudes in the linear polarization basis
$a_{0,\parallel,\perp}$ and $b_{0,\parallel,\perp}$ can
subsequently be expressed in terms of $(a,\ b,\ c)^{(\rm SM,NP)}$,
\begin{eqnarray}
a_0 & = &-a^{\rm SM}x-b^{\rm SM}(x^2-1) \, , \nonumber \\
 b_0 e^{i\phi} & = &-a^{\rm NP}x-b^{\rm NP}(x^2-1) \, , \nonumber \\
a_{\parallel} & = &\sqrt{2}a^{\rm SM} \, ,\nonumber \\
b_{\parallel}e^{i\phi} & = & \sqrt{2}a^{\rm NP} \, ,\nonumber  \\
a_{\perp} & = & c^{\rm SM}\sqrt{2(x^2-1)} \, , \nonumber \\
b_{\perp}e^{i\phi} & = & c^{\rm NP}\sqrt{2(x^2-1)} \, ,
\end{eqnarray}
where $\phi = \arg (c_{12}^{\prime})\pm \pi $.

\begin{figure}[b!]
\scalebox{0.65}{\includegraphics{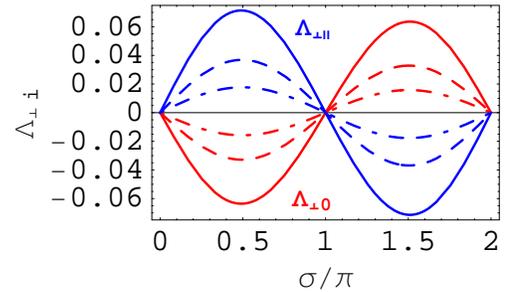}} \caption{$\L_{\perp
i}$ as a function of $\s$ for $\tilde{m}_1=200\ {\rm GeV}$ and
$\tilde{m}=2\ {\rm TeV}$. $\L_{\perp i}$ are plotted in solid,
dash and solid-dash lines for $m_{\tilde{g}}=300,\ 500,\ 800\ {\rm
GeV}$. \label{Fig_2}}
\end{figure}

In Fig. 2 we plot the observables $\L_{\perp i}$ versus the NP
weak phase $\s =\arg(c_{12}^{\prime})$ for three different values
of the gluino mass $m_{\tilde{g}}=300,\ 500,\ 800\ {\rm GeV}$,
with squark masses $\tilde{m}_1^2=(200\ {\rm GeV})^2$ and
$\tilde{m}_2^2=2\tilde{m}^2-\tilde{m}_1^2$ with $\tilde{m}=2\ {\rm
TeV}$. We note that the effects of NP can be at most a few
percent, and tend to disappear as the gluino becomes heavier
\footnote{Conservative variations of the squark scale $\tilde{m}$
and the squark mass $\tilde{m}_1$ don't produce valuable changes
in the observables $\L_{\perp i}$ from the case presented in Fig.
2.}. This can be understood from Eq.~(\ref{lambdaperpi}) together
with the expressions for
$\L_{\l\l}=a_{\l}^2+b_{\l}^2+2a_{\l}b_{\l}\cos\d_{\l}\cos\phi$. In
fact, the main contributions to $\L_{\l\l}$ are proportional to
$a_{\l}^2$ (tree level dominated) hence are of $O(1)$, while for
$\L_{\perp i}$ one has $O(b_{\l}/a_{\l})\sim O(0.01)$ at most,
being suppressed by the ratio $P^{\rm NP}/T^{\rm SM}$ with $P$ and
$T$ indicating respectively the penguin and tree terms.

We see that, for the particular model considered in this work, to
observe NP effects by performing an angular analysis for the decay
$B_d^0(t) \rightarrow J/\psi\, K^*$, one needs to be able to
extract $\L_{\perp i}$ with a precision of at least a few per
cent. On the other hand, as stressed in Ref.~\cite{LondonSinha},
no tagging or time dependent measurements are needed to measure
$\L_{\perp i}$ since it appears with the same sign in both rates
for $B_d^0(t)$ and $\bar{B}_d^0(t)$ (see Eq. (\ref{rate})).

Following from the above considerations, it is evident that the
decay $B_d^0(t) \rightarrow \phi K^*$ becomes really interesting.
This decay, contrary to  $B_d^0(t) \rightarrow J/\psi\, K^*$, is
not tree level dominated. Rather, it is of pure penguin type, and
in the model considered $a_{\l}$ and $b_{\l}$ are of the same
order. This implies now that the observables $\L_{\perp i}$ not
only differ from zero if there are NP effects, but they are
expected to be of $O(1)$, being the ratio $P^{\rm NP}/P^{\rm SM}
\sim O(1)$. Obviously for the decay $B_d^0(t) \rightarrow \phi
K^*$ the hadronic uncertainties play a more important role than
for the decay $B_d^0(t) \rightarrow J/\psi\, K^*$, plaguing the
theoretical prediction for $\L_{\perp i}$.
Furthermore, the strength of the transverse components are not yet
understood.

For $B_d^0(t) \rightarrow J/\psi\, K^*$ decay, full angular
analysis by the CLEO Collaboration \cite{CLEO} shows that the $P$
wave component is small, ${\mid P \mid^2} = {\mid A_{\perp}
\mid^2} = 0.16 \pm 0.08 \pm 0.04$, while the longitudinal
component is around $50\%$, $\G_L/\G={\mid A_0 \mid}^2= 0.52 \pm
0.07 \pm 0.04$.
%($0.65\pm 0.10\pm0.04 (CDF)$ \cite{CDF} and
%$0.97\pm0.16\pm0.15 (ARGUS)$ \cite{ARGUS}).
%
Recent measurements for the longitudinal and transverse amplitudes
have been also reported by both BaBar \cite{BaBar} and Belle
\cite{Belle} Collaborations with the respective values ${\mid
A_{\perp} \mid^2} = 0.16 \pm 0.03 \pm 0.01$, ${\mid A_0 \mid}^2=
0.60 \pm 0.03 \pm 0.02$  and ${\mid A_{\perp} \mid^2} = 0.19 \pm
0.02 \pm 0.03$, ${\mid A_0 \mid}^2= 0.62 \pm 0.02 \pm 0.03$. In
Figs. 3 and 4 we plot respectively ${\mid P \mid}^2$ and $\G_L/\G$
as a function of the NP weak phase $\s$. It seems clear that both
statistical and systematic errors need to be reduced by an order
of magnitude to discriminate a nonzero value for $\L_{\perp i}$ as
predicted by the model considered. We expect that the error on the
extracted value of $\L_{\perp i}$ will be of the same order as the
one on ${\mid P \mid}^2$ or $\G_L/\G$.

\begin{figure}[t!]
\scalebox{0.65}{\includegraphics{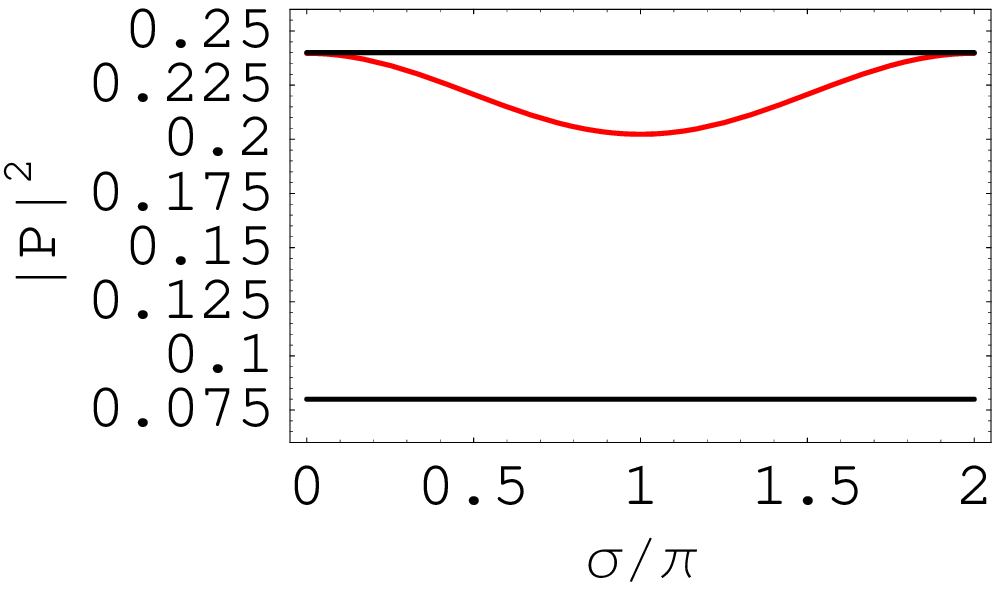}} \caption{${\mid P
\mid}^2$ as a function of $\s$ for $\tilde{m}_1=200$\ GeV,
$\tilde{m}=2\ {\rm TeV}$ and $m_{\tilde{g}}=300\ {\rm GeV}$
compared with experiment. \label{Fig_3}}
\end{figure}

\begin{figure}[t!]
\scalebox{0.65}{\includegraphics{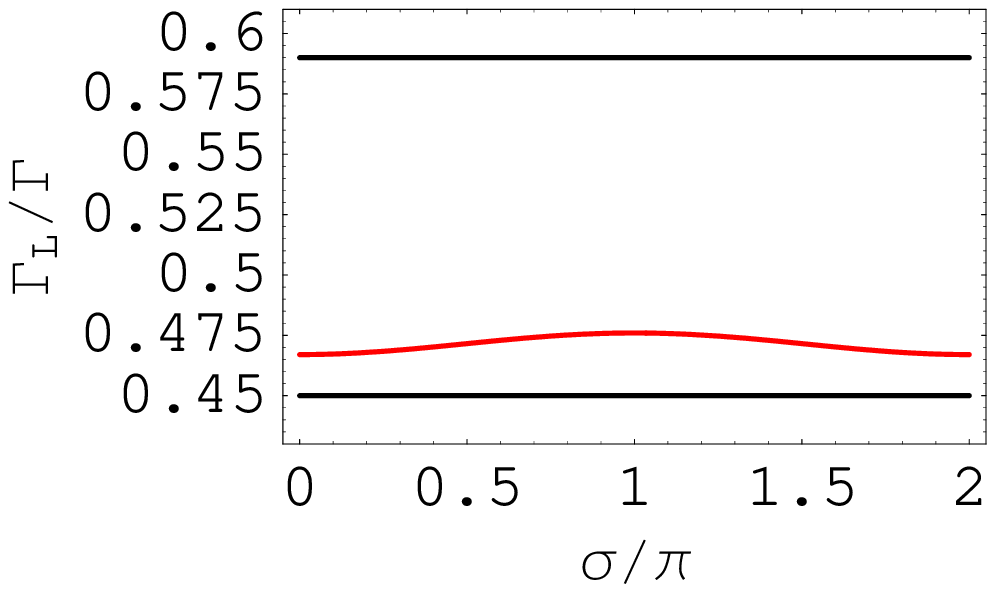}} \caption{$\G_L/\G$
as a function of $\s$ for $\tilde{m}_1=200\ {\rm GeV}$,
$\tilde{m}=2\ {\rm TeV}$ and $m_{\tilde{g}}=300\ {\rm GeV}$
compared with experiment. \label{Fig_4}}
\end{figure}

We conclude this section by presenting the results for $\sin
2\phi_{1\l}^{\rm meas} = {\rm Im}\left(
\frac{q}{p}\frac{\bar{A}_{\l}}{A_{\l}}\right)$, where $\l =
0,\parallel,\perp$ \footnote{The available measurments of $\sin
2\phi_1$ from BaBar Collaboration with value $0.22\pm 0.52$ and
Belle Collaboration with value $0.10\pm 0.45$ \cite{Moriond} are
obtained without separating final states of different $CP$. }. If
NP effects are present one will not be able to extract
$\phi_1^{\rm mix}$
\footnote{ The number of measurements for the decays $B
\rightarrow V_1 V_2$ and $\bar{B} \rightarrow \bar{V}_1 \bar{V}_2$
is fewer than the number of theoretical parameters, making it
impossible to predict $\phi_1^{\rm mix}$ purely in terms of
observables. }
(the phase of $B_d^0-\bar{B}_d^0$ mixing which in general can be
affected by NP) and the measured value of $\phi_1$, which will
depend on the helicity of the final state, will differ from the
real value of $\phi_1^{\rm mix}$ \cite{LondonSinha}. In Fig. 5 we
plot $\sin 2\phi_{1\l}^{\rm meas}$ as a function of the NP weak
phase $\s$. As for the quantities $\L_{\perp i}$, the effects of
NP on $\sin 2\phi_{1\l}^{\rm meas}$ can be at most of few per
cent. Deviations from $\sin 2\phi_{1}^{\rm mix}$ reach their
largest value at $\s = \pi/2$ and are bigger for the transverse
components, $\{ \l = \parallel, \perp \}$. On the other hand
deviations on $\sin 2\phi_{1 0}^{\rm meas}$, even if smaller,
could be easier to detect because of the higher number of
longitudinally polarized final states. On top of that by comparing
$\sin 2\phi_{1}(B\rightarrow J/\psi K_S)$ ~\cite{HeHou2} in Fig. 6
with $\sin 2\phi_{1 0}^{\rm meas}$, one can observe that
deviations from $\sin 2\phi_{1}^{\rm mix}$ have opposite signs.
\footnote{ The final state for $B\rightarrow J/\psi K_S$ is $CP$
odd on the contrary to the longitudinal component for
$B\rightarrow J/\psi K^*$ with $K^*\rightarrow K_S \pi^0$ which is
$CP$ even.}
This divergent behavior could in principle be easier to be
observed than the single deviations.

%Just for completeness we plot in Fig. 9 the ratios $r_{\l}\equiv b_{\l}/a_{\l}$.
%As pointed out in Ref. \cite{LondonSinha}, if the
%particular situation with the three strong phase differences
%$\d_{\l}$ vanishing and with $r_0=r_{\parallel}=r_{\perp}$ is realized,
%then NP effects will be hidden. In our analysis all strong phase are set to zero, but as it shown in Fig. 9 ,
%the three ratios are different, in agreement with Fig. 5  which shows NP effects.

\begin{figure}[b!]
\scalebox{0.65}{\includegraphics{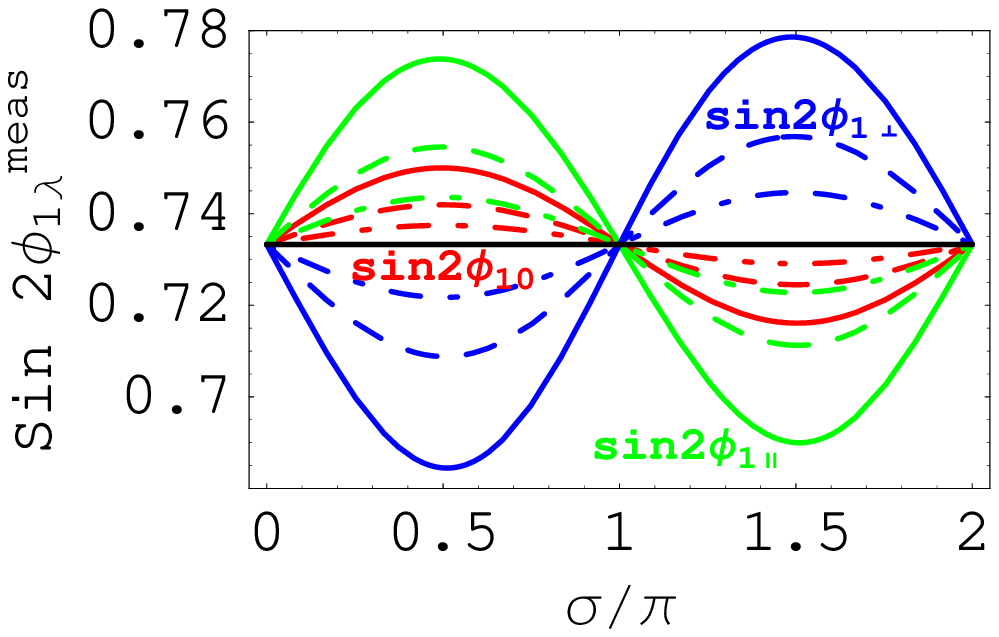}}
 \caption{$\sin 2\phi_{1\l}^{\rm meas}$ as a function of
 $\s$ for $\tilde{m}_1=200\ {\rm GeV}$ and $\tilde{m}=2\ {\rm TeV}$.
 $\sin 2\phi_{1\l}^{\rm meas}$ are plotted in solid,
 dash and solid-dash lines for $m_{\tilde{g}}=300,\ 500,\ 800\
 {\rm GeV}$. The black solid line corresponds to $\sin2\phi_1^{\rm mix}=0.733$
 \label{Fig_5}}
\end{figure}

\begin{figure}[b!]
\scalebox{0.65}{\includegraphics{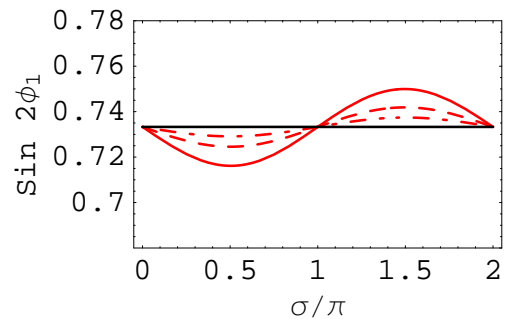}}
 \caption{$\sin 2\phi_{1}(B \rightarrow J/\psi K_S)$ as a function of
 $\s$ for $\tilde{m}_1=200\ {\rm GeV}$ and $\tilde{m}=2\ {\rm TeV}$.
 $\sin 2\phi_{1}(B \rightarrow J/\psi K_S)$ is plotted in solid, dash and solid-dash lines
 for $m_{\tilde{g}}=300,\ 500,\ 800\
 {\rm GeV}$. The black solid line corresponds to $\sin 2\phi_1^{\rm mix}=0.733$
 \label{Fig_6}}
\end{figure}

%\begin{figure}[b!]
%\scalebox{0.42}{\includegraphics{Figure_9.eps}}
%\caption{$r_{\l}$ as a function of $\s$ for $\tilde{m}_1=200 {\rm GeV}$ and $\tilde{m}=2 {\rm TeV}$.
%$r_{0}$ $(r_{\parallel})$ $[r_{\perp}]$ are plotted in red (green) [blue] solid, dash and solid-dash lines for
%$m_{\tilde{g}}=300,500,800 {\rm GeV}$.
%\label{Fig_9}}

%\end{figure}

%
%**************************************************************************
%

\section{Conclusion}

A supersymmetric extension of the SM with a light right-handed
``strange-beauty'' squark, a light gluino and a new $CP$ phase,
seems to contain all the necessary ingredients to explain the
recent $CP$ anomaly in $B_d \rightarrow \phi K_S$. In the same
framework we have calculated possible NP effects to observables
that can be extracted by the time dependent angular analysis of
$B_d \rightarrow J/\psi\, K^*$. An important role is played by the
quantities  $\L_{\perp i}$ with $i=\{0,\parallel\}$, which can be
nonzero in the presence of NP even for very small strong phase
differences. Our results show that deviations from zero can be at
most of the order of a few percent, since it is suppressed by the
ratios of the NP penguin amplitude to the SM tree amplitude. This
obviously suggests that for decays which are pure penguins, like
$B_d \rightarrow \phi K^*$, deviations from zero for the
observables $\L_{\perp i}$ are expected to be of order one.

The quantities $\sin 2\phi^{\rm meas}_{1\l}$ can also differ from
the real value $\sin 2\phi_1^{\rm mix}$ because of NP effects. In
particular $\sin 2\phi^{\rm meas}_{1 0}$ and $\sin
2\phi_{1}(B\rightarrow J/\psi K_S)$ have opposite deviations from
$\sin 2\phi_1^{\rm mix}$, and by comparing the two behaviors, NP
effects could be easier to be observed than by looking at single
deviations. As for $\L_{\perp i}$, we found that deviations are of
the order of few percent.

In conclusion, New Physics effects to $B_d \rightarrow J/\psi\,
K^*$ from the model considered in this work are found to be too
small to be observed at the current B factories.
But because of the small NP effects, $B_d^0(t) \rightarrow
J/\psi\, K^*$ %, also by virtue of its small $P$ wave component,
remains a good mode for measuring $\sin 2\phi_1^{\rm mix}$.

\acknowledgments{This work is supported in part by grants NSC
93-2112-M-002-020, NSC 93-2811-M-002-053, and NSC
93-2811-M-002-047, the BCP Topical Program of NCTS, and the MOE
CosPA project. A.S. would like to thank R. Sinha for very useful
discussions. }

%******************************************************************

\newpage

\end{document}